\documentstyle[prl,aps,preprint]{revtex}

\begin{document}
\draft
\tightenlines

\title{Observation of supercurrent enhancement in SNS junctions by
non-equilibrium injection into supercurrent carrying bound Andreev states}

\author{Jonatan Kutchinsky$^1$ \and Rafael Taboryski$^2$ \and
Claus B. S\o rensen$^3$ \and J\o rn Bindslev Hansen$^1$ \and
and Poul Erik Lindelof$^3$}

\address{$^1$Department of Physics, Technical University of Denmark, \\
Building 309, DK-2800 Lyngby, Denmark \\
$^2$Danish Institute of Fundamental Metrology, Anker Engelunds \\
Vej 1, DK-2800 Lyngby, Denmark \\
$^3$Niels Bohr Institute, University of Copenhagen, \\
Universitetsparken 5, DK-2100 Copenhagen \O , Denmark}

\date{September 28, 1999}

\maketitle

\begin{abstract}
We report for the first time enhancement of the supercurrent by means of 
injection in a mesoscopic three terminal planar SNSNS device made of Al on GaAs. 
When a current is injected from one of the superconducting Al electrodes at
an injection bias $V=\Delta(T)/e$, the DC Josephson current between the other
two superconducting electrodes has a maximum, giving evidence for an enhancement
due to a non-equilibrium injection into bound Andreev states of the underlying
semiconductor. The effect persists to temperatures where the equilibrium
supercurrent has vanished.
\end{abstract}


\noindent
The study of non-equilibrium phenomena generated by current across SN boundaries 
of SNS junctions has attracted much interest in recent years. It has been 
realised that supercurrents in SNS junctions are transmitted by the correlated 
electron-hole pairs generated by Andreev reflections at the SN boundaries. In 
this paper, we demonstrate experimentally a new supercurrent enhancement effect, 
where these correlated electron hole pairs are affected by current injection 
across a third SN boundary. 

In general, two effects have been studied in three terminal SNS devices. One is 
the field effect, where the carrier density in the normal region is controlled 
\cite{akazaki96}. The other is injection, where the normal region is connected 
to one or two other reservoirs, biased with respect to the superconductor 
\cite{schapers98,morpurgo98,baselmans99}. This modifies the quasiparticle energy 
distribution, either by simply increasing the electron temperature of the normal 
conductor, or by inducing a suitable non-thermal distribution function. These 
approaches are related to much older work \cite{mooij80}, where 
superconductivity was enhanced close to $T_c$ by redistributing quasiparticles 
by microwave irradiation or by tunnelling resulting in an enhancement of the 
energy gap and other properties derived from this.

Recent theories for SNS junctions have predicted how injection can alter the 
Josephson coupling of the junction \cite{yip98,volkov,samuelsson97}. The 
supercurrent in the junction is carried by bound Andreev states, which in
multichannel diffusive systems form a continuum. The role of a non-equilibrium 
electron distribution can be seen if one considers this spectral distribution 
$N_j$ of supercurrent carrying states \cite{yip98}, together with the energy 
distribution of electrons $f$ in the normal conductor. From these the 
supercurrent as function of the phase difference between the superconductors can 
be written:

\begin{equation}
I_{s}(\phi) = -2ev_{F} A \int_{0}^{\infty }d\varepsilon 
N_{j}(\varepsilon,\phi)\left( f(-\varepsilon)-f(\varepsilon )\right)
\label{yip} 
\end{equation}

\noindent
where $A$ is the cross-section of the conductor. The energy $\varepsilon$ is 
defined with respect to the chemical potential of the superconductors. In 
equilibrium $f$ is the Fermi distribution function. The distribution $N_j$ 
depends on the sample geometry, and can for simple systems be calculated 
directly from quasi-classical Green's function theory \cite{yip98,volkov}. In 
diffusive systems $N_j$ varies on the scale of the Thouless-energy 
$\varepsilon_{L}=\hbar D/L^{2}$, where $D$ is the diffusion constant and $L$ is 
the distance between the superconductors. For long junctions where $\Delta \gg 
\varepsilon_L$ ($\Delta$ is the superconductor energy gap), $N_j$ peaks at 
$\varepsilon\approx\varepsilon_L$. In equilibrium the critical supercurrent 
decay with the normal conductor coherence length $\xi_{N}(T)=\sqrt{\hbar D/2\pi 
k_{B}T}$. For long junctions \cite{likharev79}

\begin{equation}
I_c = I_0 \frac{\Delta^2(T)}{\Delta^2(0)} \sqrt{\frac{T_c}{T}}
\exp(-L/\xi_{N}(T))
\label{likharev}
\end{equation}

\noindent
where $I_0 = 8 \Delta^2(0) L/(e R_N \sqrt{2\pi\hbar D k_B T_c})$, and $R_N$ is 
the junction normal resistance. If however, the distribution function contains a 
sharp feature, narrower than $k_{B}T$, at the energies where the supercurrent-
carrying states are present, a very different temperature dependence and even a 
change of sign can be expected.

In the Baselmans experiment \cite{baselmans99} the non-equilibrium distribution 
function had the form of a thermally rounded step function comprised of a linear 
combination of two mutually bias-displaced Fermi functions of the normal 
reservoirs at each end of the filamentary normal conductor. The observed effect 
was however much weaker than the equilibrium supercurrent (without injection). A 
considerably stronger effect can be anticipated if the relevant non-equilibrium 
distribution function derived from the thermally rounded Fermi functions of the 
normal reservoirs is replaced by a distribution function induced between two 
voltage biased superconducting reservoirs. If the size of the normal conductor 
is smaller than the phasebreaking diffusion length, the non-equilibrium electron 
energy distribution in the normal conductor will contain replica of the sharp 
singularities in the superconducting density of states of the reservoirs. The 
fingerprints of these are seen as a sub-gap structure in the I-V characteristics 
of SNS junctions \cite{flensberg}. At a bias voltage of $V=\Delta(T)/e$ the 
singularities in the normal conductor distribution function match the Fermi 
energies of the superconductors, where the supercurrent carrying states are 
concentrated. The resulting injection induced supercurrent in the adjacent SNS 
junction is thus expected to have a maximum for an injection bias 
$V=\Delta(T)/e$, and a weak temperature dependence due to the small thermal 
smearing of the distribution function in play. Besides these non-equilibrium 
effects a broad heating effect is expected, which will always suppress the 
critical supercurrent.

In this letter we demonstrate experimentally for the first time that the 
superconductivity of an SNS-junction can be enhanced above the equilibrium value 
by injecting quasiparticles with an energy distribution which match the spectral 
supercurrent density $N_J(E)$. The effect is realised in a three terminal sample 
geometry, where three superconducting Al electrodes are connected to the same 
piece of highly doped GaAs semiconductor. One of the electrodes is used as a 
common ground for the current flow. Another electrode is used as a detector and 
the third electrode as an injector. For the particular devices studied we find 
that the injection-induced supercurrent exceeds the equilibrium supercurrent at 
temperatures above $0.6 K$ (approx.\ $\frac{1}{2}T_{c}$).

The samples were formed from a layered structure of GaAs and Al, grown in an MBE 
chamber. Here, $200 nm$ of highly doped n-GaAs were grown on an 
undoped/insulating substrate. This was then capped with 150-$200 nm$ Al. In 
order to reduce the Schottky-barrier between GaAs and Al, five layers of 
$\delta$-doping with $5\cdot 10^{13} cm^{-2}$ Si was inserted in the GaAs just 
below the Al. The Al film was subsequently deposited without breaking the 
vacuum. This resulted in a contact resistance of $8\cdot 10^{-9} \Omega cm^2$. 
Samples with a planar geometry as shown in Figure \ref{fig1} were then formed by 
removing Al in selected areas with conventional E-beam lithography and wet etch. 
Larger scale patterning to form a $20\mu m$ wide mesa structure was done with 
UV-lithography. The details of the fabrication has been published elsewhere 
\cite{taboryski96}.

The measurements were performed in a pumped $^3He$ cryostat with a base 
temperature of $235 mK$. At low temperatures the GaAs-film had a carrier density 
of $n=5.5\cdot 10^{18} cm^{-3}$ and a mean free path of $\ell=42 nm$, 
corresponding to a diffusion constant $D=127 cm^{2}/s$. In an independent weak 
localisation experiment the phase-breaking diffusion length was found to be 
$\ell_\phi\approx 5 \mu m$ at the base temperature. The Al-film had a 
superconducting critical temperature of $1.196 K$. 
All measured samples had nearly identical characteristics. The data presented in 
this paper is based on a single sample, although all samples exhibited similar 
effects. This particular sample is shown in Figure \ref{fig1}.

The three superconducting Al electrodes are connected to the same piece of GaAs 
within mesoscopic range. The distance between two neighbouring electrodes is 
approximately $400 nm$ corresponding to a Thouless-energy $\varepsilon_{L}=52 
\mu eV$. The junctions AB and BC formed individual Josephson junctions with 
normal resistances $3.3\Omega$ and $3.2\Omega$, and critical currents respectively 
$2.2\mu A$ and $3.4\mu A$ at $240 mK$, when all electrodes were kept at zero voltage. 
The junction AC was much weaker coupled because of the longer distance between A and 
C. In this experiment we define the AB junction as the detector, and the BC 
junction as the injector. The critical supercurrent was measured on the detector 
while the injector was biased at high voltages. Because the samples were 
symmetrical, the roles of the two junctions could be (and was) interchanged. 
During the experiment, a detector current was passed through electrode A, and an 
injection current was passed through electrode C, both from high impedance ($470 
k\Omega$) DC current sources. Electrode B was connected to ground through an 
Ampere-meter, in order to check that no current leaks were present in the set-
up. The detector (AB) and injector (BC) voltages were measured by separate leads 
to the superconductors. All leads to the sample were fitted with $50 cm$ low 
temperature THERMOCOAX filters providing high frequency power attenuation of 
$12.5\cdot \sqrt{f[GHz]} dB$ and room temperature $\pi$ filters giving 
approximately $20 dB$ attenuation at $700 kHz$.

Both the detector and injector junctions showed a ``Fraunhofer´´-type of pattern 
with up to 10 well-developed lobes in the critical supercurrent vs.\ applied 
magnetic field, indicating a high degree of homogeneity of the junctions. Before 
each measurement series on the 3-terminal samples, the magnetic field was 
carefully zeroed to within a small fraction of a flux quantum \cite{footmagnetic}.

In Figure \ref{fig2} we show the detector supercurrent at various injection 
currents. In the left panel, the detector critical current $I_{c}$ is plotted as 
function of the injection current at three different temperatures. Above the 
highest injection currents shown in the figure, the central electrode suddenly 
went normal due to a current in excess of its critical current and above this, 
injector bias measurements could not be made. Each data point is the result of a 
fit to a noise rounded resistively shunted junction (RSJ) model 
\cite{ambageokar} of individual I-V characteristics of the detector junction. In 
the right panel of Figure \ref{fig2} we show three examples of the fits. The 
injection currents and fitted $I_c$ of these are indicated with arrows in the 
left panel. The fitted noise temperature varied randomly in the range
$1-3 K$ due to the error on the measured data \cite{footrsj}. 
The I-V characteristics of the detector junction show current offsets. The 
offsets are simply a fraction of the injector current and are exclusively due to 
the sample geometry, where the three electrodes are connected to the same piece of 
GaAs. Consequently the detector current $I_{det}$ has to be compensated by about 
one fourth of the injection current for the detector to be at zero bias. 

In the left panel of Figure \ref{fig2} we see, that the critical supercurrent 
exhibit two clear features as function of injection current (and voltage). The 
equilibrium supercurrent appears to be falling off rapidly with increasing 
injection current, and is seen to be strongly temperature dependent. The next 
maximum appearing for higher injection currents is interpreted as a non-equilibrium 
supercurrent. At very high injection currents, close to the critical current of 
the central electrode, we observe a weak enhancement of the critical current. 
This effect remains unexplained. The non-equilibrium critical current builds up 
with increasing injection current to a maximum (at $400 
mK$ indicated by C in Figure \ref{fig2}). We see that while the equilibrium 
supercurrent is dominating at $400 mK$, at $900 mK$ it has almost vanished. The 
non-equilibrium supercurrent on the other hand is almost temperature independent 
\cite{foottemp} and by far exceeds the equilibrium supercurrent at $900 mK$.
At $600 mK$ the two 
critical currents have equal magnitude. Although, in the figure we mostly plot 
the properties for positive injection currents, we would like to emphasise that 
the effect was completely symmetrical upon reversal of current direction. The 
effects were reproduced in several samples.

In Figure \ref{fig3} we show a detailed plot of the temperature dependence of 
both the critical current at zero injection, and the optimal non-equilibrium 
current. The detector $I_{c}$ without injection (squares) has a strong 
temperature dependence, which has been fitted to Eq. (\ref{likharev}) with 
parameters $I_0 = 4.9\mu A$ and $\xi_N(T_c) = 115 nm$. From the independently 
measured sample parameters we find $I_0 = 363 \mu A$ and $\xi_N(T_c) = 113 nm$.
The large deviation of the 
prefactor $I_0$ may be attributed to the fact that the calculation of Eq.\ 
(\ref{likharev}) does not take interface barriers into account. The other curve 
(triangles) shows the maxima of the injection induced non-equilibrium $I_{c}$ as 
exemplified for three temperatures in Figure \ref{fig2}. This only decreases 
slightly with increasing temperature. In the insert, we plot the corresponding 
voltage across the injector junction for each of these points. The injector 
voltage varied slightly during each detector I-V measurement. The plotted value 
was taken where the detector current $I_{det}$ matched the earlier mentioned 
current offset. The full curve in the insert shows the BCS theory gap function, 
where $T_{c}=1.196K$ has been measured, and $\Delta (0)=158\mu eV$ has been 
adjusted from the bulk value $175\mu eV$ to fit the data. It is clearly seen 
that the injection bias giving the maximum non-equilibrium supercurrent, 
corresponds to the superconductor energy gap $\Delta(T)$. Together with the weak 
temperature dependence of the non-equilibrium critical current, we take this as 
the most important experimental indication that the observed injection induced 
phenomena are related to a non-equilibrium population of the supercurrent 
carrying states. This non-equilibrium population is induced by Andreev 
reflections on the superconducting injection electrode, and is possessing a 
sharp kink at the superconducting gap energy. This kink originates from the 
singularities in the density of states of the superconductor. The functional 
dependence of the spectral supercurrent density of Eq.\ (\ref{yip}) has not been 
calculated for the planar geometry of the samples used in the experiment, 
however the form of the distribution function may allow us to probe the varying 
part of $N_j$. We believe the observed phenomena are related to the earlier 
observations published in Ref.\ \cite{kutchinsky97}, where a maximum in 
oscillation amplitude in a flux sensitive interferometer was observed for the 
bias voltages $V=\Delta/e \pm \varepsilon_L$. Here the phase dependent 
(coherent) change of the normal resistance was probed instead of the 
supercurrent.

In conclusion, we have measured the injection-induced supercurrent in a three 
terminal device. Above $600 mK$ the induced non-equilibrium supercurrent exceeds 
the equilibrium (zero injection) supercurrent. The optimal injection induced 
critical current is observed when the injector electrode is biased at 
$V=\Delta(T)/e$. Further studies, are needed to map out the detailed properties
of the non-equilibrium supercurrent, e.g. the magnetic field dependence and the
d.\ c.\ Josephson effects.

We acknowledge useful discussions with Vitaly Shumeiko and Jonn Lantz. We thank 
the III-V Nanolab of the Niels Bohr Institute for providing the processing 
facilities. This work has been supported by the Danish Technical Research 
Council, and the Velux Foundation.


\begin{figure}
\caption{%
The three terminal device and the measurement set-up. Top-panel:
Optical photo of the sample, with SEM\ close-up. Both seen from above. The
bright areas consist of Al, the darker GaAs. The SEM picture shows the
roughness of the Al-edge, due to the crystal grain boundaries. Bottom:\
Schematic cut-through of the sample, with measurement leads indicated. The
injector and detector currents were controlled by high-impedance current
sources. The middle lead (output) was connected to ground. The output
current $I_{\text{out}}$ was measured to make sure there were no leaks or
nonlinearities in the set-up. The voltages $V_{\text{inj}}$ and $V_{\text{det}}$
were measured through separate leads to the superconducting electrodes.}
\label{fig1}
\end{figure}

\begin{figure}
\caption{%
The detector supercurrent at various injection currents. Left
panel: The detector critical currents $I_{c}$ plotted as function of the
injection current at three different temperatures. The data for $T = 600 mK$ and 
$400 mK$ has been shifted upwards with $0.1\mu A$ and $0.2 \mu A$ respectively. 
Right panel: The $I/V$-characteristics of the detector at $400mK$ and (A) no 
injection, (B) $18\mu A$ injection, and (C) $64\mu A$ injection. Circles 
represent measurements. The curves represent fits to the RSJ-model. The three 
plots have separate y-axes indicated by the arrows.}
\label{fig2}
\end{figure}

\begin{figure}
\caption{%
The critical supercurrent $I_{c}$ of the detector junction as function of temperature.
Squares: Ic with no injection. Triangles: non-equilibrium supercurrent Ic at the
optimal injection current for a given temperature. The curve through the
no-injection points is a fit with Eq.\ (\ref{likharev}). To guide the eye, lines 
connect the optimal injection data points. Inset: Circles indicate the optimal 
injector voltage for a given temperature. The curve shows the BCS theory gap 
function.}
\label{fig3}
\end{figure}



\begin{references}
\bibitem{akazaki96}T. Akazaki, H. Takayanagi, J. Nitta, and T. Enoki, Appl. 
Phys. Lett. {\bf 68}, 418 (1996)

\bibitem{schapers98}Th. Sch\"{a}pers et. al., Appl. Phys. Lett. {\bf 73}, 2348 
(1998)

\bibitem{morpurgo98}A. F. Morpurgo, T. M. Klapwijk, and B. J. van Wees, 
Appl. Phys. Lett. {\bf 72}, 966 (1998)

\bibitem{baselmans99}J. J. A. Baselmans, A. F. Morpurgo, B. J. van Wees, and T. 
M. Klapwijk, Nature {\bf 397}, 43 (1999)

\bibitem{mooij80}For a review see J. E. Mooij, in {\it Nonequilibrium 
Superconductivity, 
Phonons, and Kapitza Boundaries}, ed. by K. E. Gray (Plenum, New York, 
1981)

\bibitem{yip98}S.-K. Yip, Phys. Rev. B {\bf 58}, 5803 (1998);
F.Wilhelm, G.Sch\"{o}n, A.D.Zaikin, Phys. Rev. Lett. {\bf 81}, 1682 (1998)

\bibitem{volkov}A. F. Volkov and H. Takayanagi, Phys. Rev. B {\bf 56},  11184 
(1997); A. F. Volkov and V. V. Pavlovskii, JETP Lett. {\bf 64}, 670 (1996)

\bibitem{samuelsson97}P. Samuelsson, V. S. Shumeiko, and G. Wendin, Phys. Rev. B 
{\bf 56}, 5763 (1997); P. Samuelsson, J. Lantz, V. S. Shumeiko, and G. Wendin, 
cond-mat/9904276

\bibitem{likharev79}K. K. Likharev, Rev. Mod. Phys. {\bf 51}, 101 (1979)

\bibitem{flensberg}M. Octavio, M. Tinkham, G. E. Blonder, and T. M. Klapwijk, 
Phys. Rev. B {\bf 27}, 6739 (1983); K. Flensberg, J. Bindslev Hansen, and M. 
Octavio, Phys. Rev. B {\bf 38}, 8707 (1988).

\bibitem{taboryski96}R. Taboryski et. al., Appl. Phys. Lett. {\bf 69}, 656 
(1996)

\bibitem{footmagnetic}The magnetic field created by the injection current is
negligible in comparison with one flux quantum through the junction area.

\bibitem{ambageokar}V. Ambageokar and B. I. Halperin, Phys. Rev. Lett. {\bf 22}, 
1364 (1969)

\bibitem{footrsj}The RSJ model assumes white noise. Our experimental setup is
however dominated by noise below 1 kHz. The extracted noise temperature is therefore
not the electron temperature in the device, but rather represents external
low-frequency noise.

\bibitem{foottemp}The highly non-equilibrium condition of the experiment,
makes it difficult to define an electron temperature in the active normal
conducting region. Since the inelastic scattering length is much longer
than this region, the thermalisation of the electrons mainly take place
in the superconducting electrodes, far from the junction compared with the
lengthscales relevant for Andreev scattering.

\bibitem{kutchinsky97}J. Kutchinsky et. al., Phys. Rev. Lett. {\bf 78}, 931 
(1997); Phys. Rev. B {\bf 56}, 2932 (1997); R. Taboryski et. al., Superlattices 
and Microstructures {\bf 25}, 829 (1999)

\end{references}
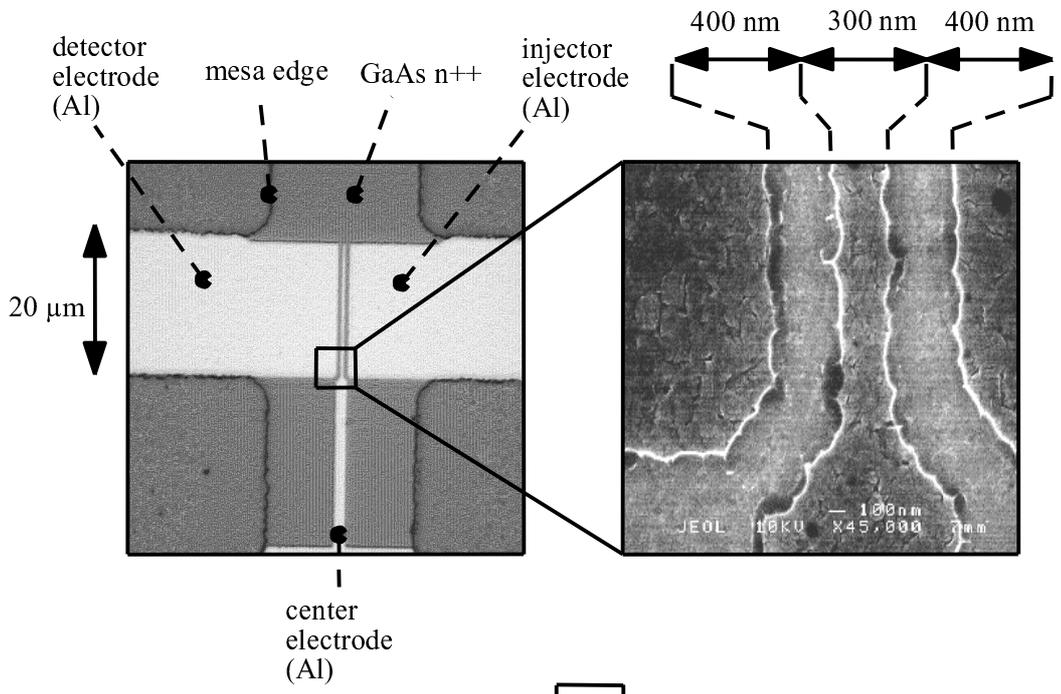
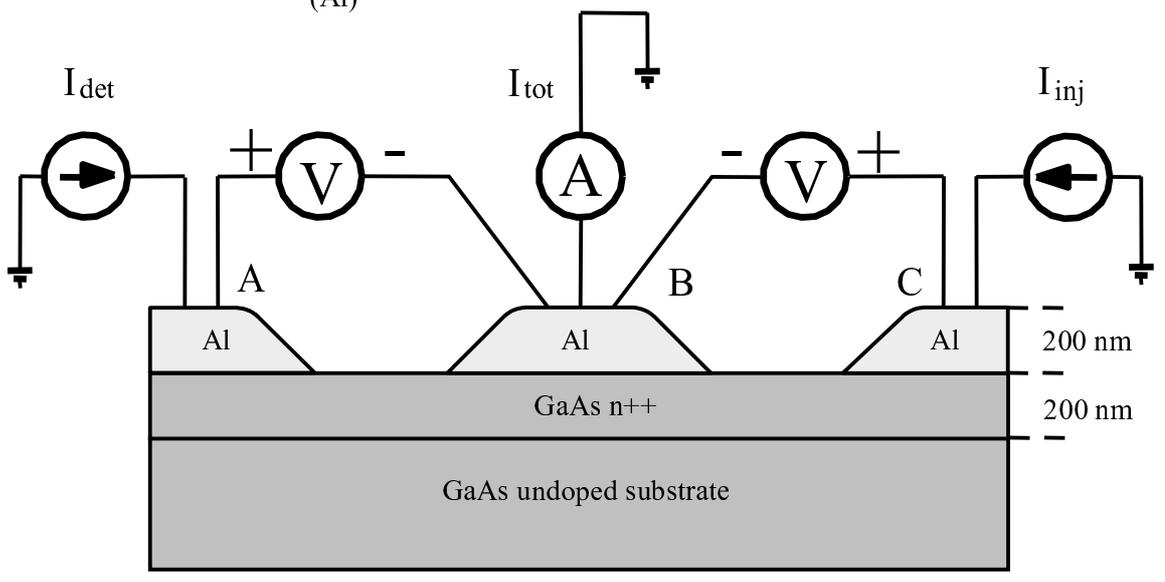
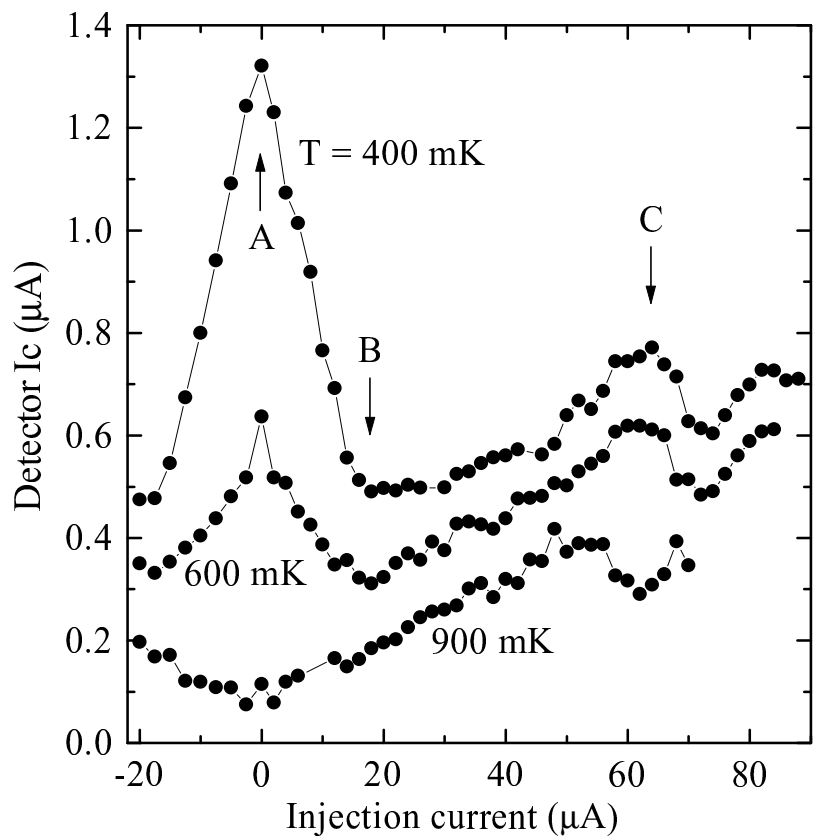
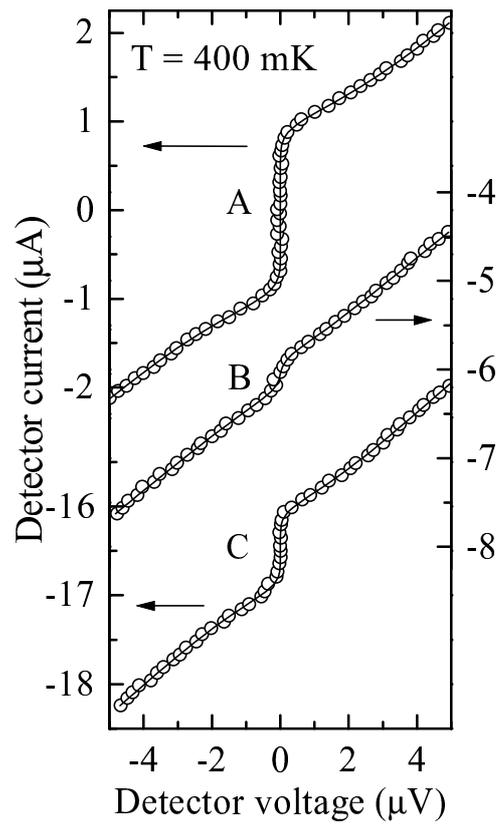
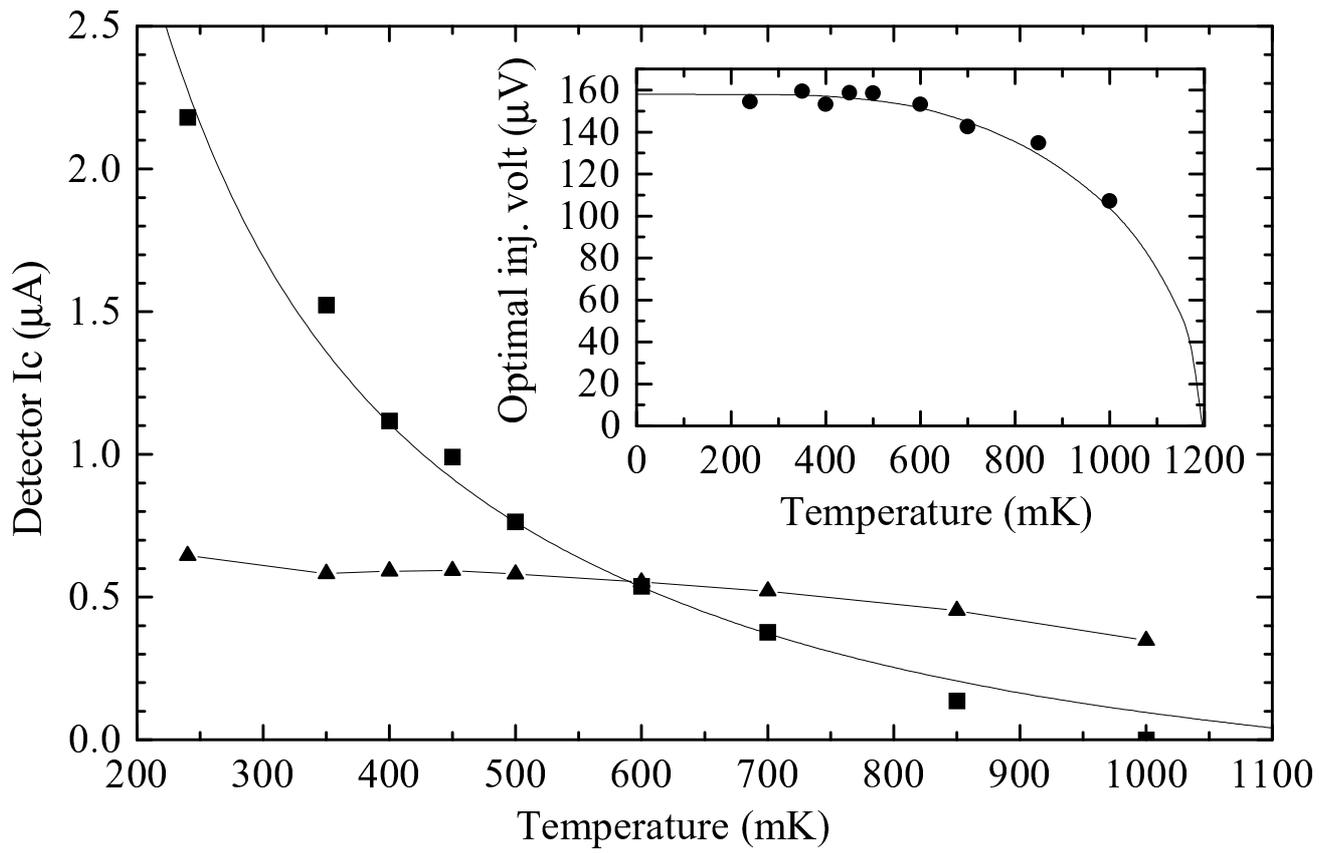
\end{document}